\documentclass[a4paper]{jpconf}
\usepackage{graphicx}% Include figure files
\usepackage{amssymb}
\usepackage{bm}% bold math

\newcommand{\chg}{\ensuremath{e}}
\newcommand{\bmag}{\ensuremath{\mu_B}}
\newcommand{\mgm}{\ensuremath{\mu}}
\newcommand{\elm}{\ensuremath{\epsilon}}
\newcommand{\afl}{\ensuremath{\ell}}

\begin{document}
\title{Neutrino-atom collisions}
\author{Konstantin A Kouzakov$^{1}$ and Alexander I Studenikin$^{2,3}$}
\address{$^1$ Department of Nuclear Physics and Quantum Theory of Collisions, Faculty of Physics, Lomonosov Moscow State University, Moscow 119991, Russia}%Lines break automatically or can be forced with \\
% \ead{kouzakov@srd.sinp.msu.ru}
%

%
\address{$^{2}$ Department of Theoretical Physics, Faculty of Physics, Lomonosov Moscow State University, Moscow 119991,
Russia\\
$^{3}$ Joint Institute for Nuclear Research, Dubna 141980, Moscow
Region,
Russia}%Lines break automatically or can be forced with \\
%

%
% \ead{studenik@srd.sinp.msu.ru}
%

\ead{kouzakov@srd.sinp.msu.ru}

\begin{abstract}
Neutrino-atom scattering provides a sensitive tool for probing nonstandard interactions of massive neutrinos in laboratory measurements. The ionization channel of this collision process plays an important role in experiments searching for neutrino magnetic moments. We discuss some theoretical aspects of atomic ionization by massive neutrinos. We also outline possible manifestations of neutrino electromagnetic properties in coherent elastic neutrino-nucleus scattering.    
%The current theoretical studies on neutrino-atom collisions are briefly overviewed. The ionization channel of these processes, which is studied in experiments searching for neutrino magnetic moments, is brought into focus. Recent developments in the theory of atomic ionization by impact of reactor antineutrinos are discussed. It is shown that the stepping approximation is
%well applicable for the data analysis practically down to the ionization threshold. The potential effects of neutrino electromagnetic properties on coherent elastic neutrino-nucleus scattering are also discussed.
%
\end{abstract}
\section{Introduction}
\label{intro}
The neutrino oscillations determined by many dedicated experiments (see the review articles
\cite{Beringer:1900zz,Giunti:2007ry,Bilenky:2010zza,Xing:2011zza,GonzalezGarcia:2012sz,Bellini:2013wra}) are generated by neutrino masses and mixing~\cite{Pontecorvo:1957cp,Pontecorvo:1957qd,Maki:1962mu,Pontecorvo1968}.
Therefore, the Standard Model (SM) must be extended to account for the
neutrino masses. Various extensions of the
SM predict different properties for neutrinos~\cite{Xing:2011zza,Ramond:1999vh,Mohapatra:2004}. In many such extensions, neutrinos
acquire also electromagnetic properties through quantum loops'
effects, thus allowing interactions of neutrinos with
electromagnetic fields and electromagnetic interactions of
neutrinos with charged particles (see~\cite{Giunti_RMP2015} for the most comprehensive review of neutrino electromagnetic properties and interactions). 
%Hence, the theoretical and experimental study of neutrino
%electromagnetic interactions is a powerful tool in the search for
%a more fundamental theory beyond the Standard Model.

The most well
studied and understood among the neutrino electromagnetic
characteristics are the dipole magnetic and electric moments.
The diagonal magnetic and electric moments of a Dirac neutrino in
the minimally-extended SM with right-handed neutrinos,
derived for the first time in \cite{Fujikawa:1980yx}, are
respectively
\begin{equation}
\label{mu_D}
    \mgm_{ii}
  = \frac{3eG_F m_i}{8\sqrt{2}\pi^2}\approx 3.2\times 10^{-19}
  \mu_B\left(\frac{m_i}{1 \, {\rm eV}}\right), \qquad \elm_{ii}=0,
  \end{equation}
where $m_i$ is the neutrino mass and $\mu_B$ is the Bohr magneton. According to (\ref{mu_D}), the
value of the neutrino magnetic moment is very small. However, in
many other theoretical frameworks (beyond the minimally-extended
SM) the neutrino magnetic moment can reach values that
are of interest for the next generation of terrestrial experiments
and also accessible for astrophysical observations. The current
best laboratory upper limit on a neutrino magnetic moment,
$\mgm_\nu \leq 2.9 \times 10^{-11} \mu_B$ (90\% CL), has been obtained by
the GEMMA collaboration \cite{Beda:2012zz}. The best
astrophysical limit, $\mgm_\nu\leq 3 \times 10^{-12} \mgm_B$ (90\% CL)
\cite{Raffelt:1990pj}, comes from the constraints on the possible delay of helium ignition of a red giant star in globular clusters due to the cooling induced by the energy loss in the plasmon-decay process $\gamma^*\to\nu\bar{\nu}$. 

The most sensitive and
widely used method for the experimental investigation of the
neutrino magnetic moment is provided by direct laboratory
measurements of low-energy elastic scattering of neutrinos and
antineutrinos with electrons in reactor, accelerator and solar
experiments. Detailed descriptions of these experiments can be found in~\cite{Giunti_RMP2015,Wong:2005pa, Balantekin:2006sw,
Beda:2007hf,Giunti:2008ve,Broggini:2012df}.
%
%Extensive experimental studies of the neutrino magnetic moment,
%performed during many years, are stimulated by the hope to observe
%a value much larger than the prediction in Eq.~(\ref{mu_D}) of the
%minimally extended Standard Model with right-handed neutrinos. It
%would be a clear indication of new physics beyond the extended
%Standard Model. For example, the effective magnetic moment in
%$\bar\nu_{e}$-$e$ elastic scattering in a class of extra-dimension
%models can be as large as about $10^{-10} \bmag$
%\cite{Mohapatra:2004ce}. Future higher precision reactor
%experiments can therefore be used to provide new constraints on
%large extra-dimensions.
%
The possibility for neutrino-electron elastic scattering due to
the neutrino magnetic moment was first considered in~\cite{Carlson:1932rk} and the cross section of this process was
calculated in~\cite{Bethe:1935cp} (for related short historical
notes see~\cite{Kyuldjiev:1984kz}). In~\cite{Domogatsky:1971tu} the cross section of~\cite{Bethe:1935cp} was corrected and the antineutrino-electron
cross section was considered in the context of the earlier
experiments with reactor antineutrinos of~\cite{Cowan:1954pq,Cowan:1957pp}, which were aimed to reveal the
effects of the neutrino magnetic moment. Discussions on the
derivation of the cross section and on the optimal conditions for
bounding the neutrino magnetic moment, as well as a collection of
cross section formulae for elastic scattering of neutrinos
(antineutrinos) on electrons, nucleons, and nuclei can be found in
\cite{Kyuldjiev:1984kz,Vogel:1989iv}.

In the above-mentioned experiments, the electrons are bound into atoms in the employed detectors and, hence, the elastic scattering of neutrinos and antineutrinos on these electrons can induce atomic ionization (see the review article~\cite{Kouzakov:2014ahe} and references therein). With lowering the energy-transfer value𝑇
an additional collision channel apart from ionization opens
up, namely, the coherent elastic neutrino-nucleus scattering~\cite{Freedman:1973yd}. This particular channel has not been
experimentally observed so far, but it is expected to be
accessible, for example, in the reactor experiments when lowering the energy
threshold of the employed Ge
detectors down to several hundred eV~\cite{Wong:2011zzb,Li:2013fla,Li:2013ewa}. Any deviation of
the measured cross section from the very precisely known SM value~\cite{Drukier:1984} will provide a signature of the physics beyond the SM. Therefore, it is important to examine how the neutrino electromagnetic interactions can contribute to the coherent elastic neutrino-nucleus
scattering . 

The paper is organized as follows. Section~\ref{nu-e} is devoted to scattering of neutrinos on free and atomic electrons due to neutrino magnetic moments. The role of the center-of-mass atomic motion in the processes of atomic ionization by neutrinos is also discussed. In section~\ref{nu-N}, we analyze how the neutrino magnetic moment, millicharge and charge radius can manifest themselves in neutrino-nuclues coherent scattering. 
\section{Neutrino-electron elastic scattering}
\label{nu-e}
Let us consider the process
\begin{equation}
\nu_{\afl} + e^{-} \to \nu_{\afl'} + e^{-}, \label{D035}
\end{equation}
where a neutrino with flavor $\afl=e,\mu,\tau$ and energy $E_{\nu}$ elastically scatters off a free electron (FE) at rest in the laboratory frame. Due to neutrino mixing, the final neutrino flavor $\afl'$ can be different from $\afl$.
There are two observables: the kinetic energy $T_{e}$ of the
recoil electron and the recoil angle $\chi$ with respect to the
neutrino beam, which are related by
\begin{equation}\label{D036}
\cos\chi =
\frac{E_{\nu}+m_{e}}{E_{\nu}}\Big[\frac{T_{e}}{T_{e}+2m_{e}}\Big]^{1/2}
.
\end{equation}
The electron kinetic energy is constrained from the
energy-momentum conservation by
\begin{equation}
T_{e} \leq \frac {2E_{\nu}^{2}}{2E_{\nu} + m_{e}} .
\end{equation}

Since, in the ultrarelativistic limit, the neutrino magnetic
moment interaction changes the neutrino helicity and the SM weak interaction conserves the neutrino helicity, the two
contributions add incoherently in the cross section,
which can be written as~\cite{Vogel:1989iv}
\begin{equation}\label{D037}
\frac{d\sigma_{\nu_{\afl}e^{-}}}{dT_{e}} =
\left(\frac{d\sigma_{\nu_{\afl}e^{-}}}{dT_{e}}\right)_{\rm SM}^{\rm FE}
+
\left(\frac{d\sigma_{\nu_{\afl}e^{-}}}{dT_{e}}\right)_{\rm mag}^{\rm FE}
.
\end{equation}

The weak-interaction cross section is given by
\begin{eqnarray}
\left(\frac{d\sigma_{\nu_{\afl}e^{-}}}{dT_{e}}\right)_{\rm SM}^{\rm FE}
=  \frac{G^{2}_F m_{e}}{2\pi} \bigg\{
(g_{V}^{\nu_{\afl}} + g_{A}^{\nu_{\afl}})^{2} %\nonumber
%\\
%\null & \null 
+ (g_{V}^{\nu_{\afl}} - g_{A}^{\nu_{\afl}})^{2}
\left(1-\frac{T_{e}}{E_{\nu}}\right)^{2} 
%\nonumber
%\\
%\null & \null 
+ \left[ (g_{A}^{\nu_{\afl}})^{2} -
(g_{V}^{\nu_{\afl}})^{2} \right] \frac{m_{e}T_{e}}{E_{\nu}^{2}}
\bigg\} , \label{D038}
\end{eqnarray}
with the standard coupling constants $g_{V}$ and $g_{A}$ given by
\begin{eqnarray}
g_{V}^{\nu_{e}} = 2\sin^{2} \theta_{W} + 1/2 ,\qquad g_{V}^{\nu_{\mu,\tau}} = 2\sin^{2} \theta_{W} - 1/2
, \qquad
 g_{A}^{\nu_{e}} = 1/2, \qquad g_{A}^{\nu_{\mu,\tau}} = - 1/2.\label{D039}
%\\
%g_{V}^{\nu_{\mu,\tau}} = 2\sin^{2} \theta_{W} - 1/2
%, \qquad g_{A}^{\nu_{\mu,\tau}} = - 1/2 .
%\label{D040}
\end{eqnarray}
For antineutrinos one must substitute $g_A \to -g_A$.

The neutrino magnetic-moment contribution to the cross section is
given by~\cite{Vogel:1989iv}
\begin{equation}
\left(\frac{d\sigma_{\nu_\afl e^-}}{dT_e}\right)_{\rm mag}^{\rm FE}
= \frac{\pi\alpha^2}{m_e^2}\left(\frac{1}{T_e}-\frac{1}{E_\nu}\right)
\left(\frac{\mgm_{\nu_\ell}}{\bmag}\right)^2 , \label{D041}
\end{equation}
where $\mgm_{\nu_\afl}$ is the effective magnetic moment~\cite{Giunti_RMP2015}. It is traditionally called
``magnetic moment'', but it receives 
contributions from both the electric and magnetic dipole moments.

The two terms $(d\sigma_{\nu_\afl e^-}/dT_e)_{\rm SM}^{\rm FE}$ and
$(d\sigma_{\nu_\afl e^-}/dT_e)_{\rm mag}^{\rm FE}$ exhibit quite
different dependencies on the experimentally observable electron
kinetic energy $T_{e}$.
%as illustrated in Fig.~\ref{D042} taken from
%\cite{Balantekin:2013sda} (see also
%\cite{Vogel:1989iv,Beda:2007hf}).
One can see that small values of the neutrino magnetic moment can
be probed by lowering the electron recoil-energy threshold. In
fact, considering $T_{e} \ll E_{\nu}$ in formulas~(\ref{D038}) and~(\ref{D041}), one can find that
$(d\sigma/dT_{e})_{\rm mag}^{\rm FE}$ exceeds
$(d\sigma/dT_{e})_{\rm SM}^{\rm FE}$ for
\begin{equation}
\label{nu-e_mag_SM}
T_{e} \lesssim \frac{\pi^2\alpha^2}{G_F^2m_e^3}
\left(\frac{\mgm_{\nu_\afl}}{\bmag}\right)^2.
\end{equation}

The current experiments with reactor antineutrinos have reached
threshold values of $T_e$ as low as few keV. These experiments are likely to
further improve the sensitivity to low energy deposition in the
detector. At low energies, however, one can expect a modification of
the free-electron formulas~(\ref{D038}) and~(\ref{D041}) due
to the binding of electrons in the germanium atoms, where, e.g., the
energy of the $K_\alpha$ line, 9.89\,keV, indicates that at least
some of the atomic binding energies are comparable to the already
relevant to the experiment values of $T_e$. It was
demonstrated~\cite{Kouzakov:2010tx,Kouzakov:2011ig,Kouzakov:2011ka,Kouzakov:2011vx,Kouzakov:2011uq}
by means of analytical and numerical calculations that the atomic-binding effects are adequately described by the so-called stepping
approximation introduced in~\cite{kopeikin97} from interpretation
of numerical data. According to
the stepping approach, %the SM and NMM contributions are simply given by
\begin{eqnarray}
\left(\frac{d\sigma_{\nu_{\afl}e^{-}}}{dT_{e}}\right)_{\rm SM}=\left(\frac{d\sigma_{\nu_{\afl}e^{-}}}{dT_{e}}\right)_{\rm SM}^{\rm FE}\sum_j n_j\theta(T_e-I_j), \label{d1sw}\\
\left(\frac{d\sigma_{\nu_{\afl}e^{-}}}{dT_{e}}\right)_{\rm mag}=\left(\frac{d\sigma_{\nu_{\afl}e^{-}}}{dT_{e}}\right)_{\rm mag}^{\rm 
FE}\sum_j n_j\theta(T_e-I_j),\label{d1s}
\end{eqnarray}
where the $j$ sum runs over all occupied atomic sublevels, with
$n_j$ and $I_j$ being their occupations and ionization energies. Numerical calculations~\cite{Kouzakov:2014pepanlett,Chen:2014plb} beyond the model of independent atomic electrons exhibit suppression of atomic factors relative to the stepping approximation when the energy-transfer value is close to the ionization threshold. This suppression can be explained by the electron-correlation effects~\cite{Kouzakov:2014ahe}.
%The following
%important conclusions can be drawn from the stepping
%approximation~(\ref{step}). Firstly, the atomic effects reduce the SM and
%NMM contributions compared to their free-electron values. Secondly, the
%ratio between the SM and NMM contributions is not affected by the atomic
%binding effects.

As shown in~\cite{Kouzakov:2016ichep}, the cross sections~(\ref{d1sw}) and~(\ref{d1s}) become suppressed and even
vanish when $T_e\to I_j$ due to atomic recoil. The following
estimate for the energy range where the atomic-recoil effects are
important was derived within the Thomas-Fermi model:
$$
T_e-I\lesssim2Z^{4/3}E_h\frac{m_e}{M_N},
$$
where $E_h=\alpha^2m_e=27.2$\,eV is the Hartree energy and $M_N$ is the nuclear mass. For germanium
($Z=32$) one obtains $T_e-I\lesssim0.04$\,eV. This energy scale appears to be
insignificant for the experiments searching for magnetic moments of
reactor antineutrinos~\cite{Beda:2012zz,Wong:2005pa}. 
%These experiments have reached threshold values of $T$ as low as few keV,
%which are already below the ionization threshold for $K$ electrons
%in germanium ($I_K\approx11$\,keV). These atomic electrons are most
%strongly bound, and the Thomas-Fermi average $p$ value substantially
%underestimates their average momentum. The latter can be evaluated using
%the virial theorem as $p=\sqrt{2m_eI_K}$. The corresponding energy scale
%is $T-I\lesssim0.3$\,eV. It appears to be about an order of magnitude
%larger than that given by the Thomas-Fermi model, but it is still
%insignificant for the discussed experiments.

%
\section{Neutrino-nucleus coherent scattering}
\label{nu-N} 
%
%As mentioned above, the most
%sensitive probe of neutrino electromagnetic properties is provided
%by direct laboratory measurements of (anti)neutrino-electron
%scattering at low energies in solar, accelerator and reactor
%experiments (their detailed description can be found
%in~\cite{Wong:2005pa, Balantekin:2006sw,
%Beda:2007hf,Giunti:2008ve, Broggini:2012df,Giunti:2014ixa}). The coherent elastic neutrino-nucleus
%scattering~\cite{Freedman:1973yd} has not been
%experimentally observed so far, but it is expected to be
%accessible in the reactor experiments when lowering the energy
%threshold of the employed Ge
%detectors~\cite{Wong:2011zzb,Li:2013fla,Li:2013ewa}.
%
Let us consider the case when an electron neutrino scatters on a
spin-zero nucleus with even numbers of protons and neutrons, $Z$
and $N$. The matrix element of this process, taking into account
the neutrino electromagnetic properties, reads~\cite{Kouzakov:2015nppp}
\begin{eqnarray}
\label{M}
\mathcal{M}&=&\left[\frac{G_F}{\sqrt{2}}\bar{u}(k^\prime)\gamma^\mu(1-\gamma_5)u(k)C_V
+\frac{4\pi Ze}{q^2}\left({\chg_{\nu_e}}+\frac{e}{6}\,q^2\langle r_{\nu_e}^2\rangle\right)\bar{u}(k^\prime)\gamma^\mu u(k)
\right.\nonumber\\&{}&\left.-\frac{4\pi
Ze\mgm_{\nu_e}}{q^2}\bar{u}(k^\prime)\sigma^{\mu\nu}q_\nu
u(k)\right]j_\mu^{(N)},
\end{eqnarray}
where $C_V=[Z(1-4\sin^2\theta_W)-N]/2$,
$j_\mu^{(N)}=(p_\mu+p_\mu^\prime)F(q^2)$, with $p$ and $p^\prime$
being the initial and final nuclear four-momenta, $\chg_{\nu_e}$ and $\langle r_{\nu_e}^2\rangle$ are the neutrino millicharge and charge radius~\cite{Giunti_RMP2015}. For neutrinos
with energies of a few MeV the maximum momentum transfer squared
($|q^2|_{\rm max}=4E_\nu^2$) is still small compared to $1/R^2$,
where $R$, the nucleus radius, is of the order of
$10^{-2}-10^{-1}$\,MeV$^{-1}$. Therefore, the nuclear elastic form
factor $F(q^2)$ can be set equal to one. Using (\ref{M}), we
obtain the differential cross section in the nuclear-recoil energy transfer
$T_N$ as a sum of two components. The first
component conserves the neutrino helicity and can be presented in
the form
\begin{equation}
\label{h.-c.}
\left(\frac{d\sigma_{\nu_{e}N}}{dT_{N}}\right)_{\rm SM}^{\rm Q}=\eta^2\,\left(\frac{d\sigma_{\nu_{e}N}}{dT_{N}}\right)_{\rm SM},
\qquad \eta=1-\frac{\sqrt{2}\pi
eZ}{G_FC_V}\left[\frac{\chg_{\nu_e}}{M_NT_N}-\frac{e}{3}\,\langle
r_{\nu_e}^2\rangle\right],
\end{equation}
where $M_N$ is the nuclear mass, and
\begin{equation}
\label{SM}
\left(\frac{d\sigma_{\nu_{e}N}}{dT_{N}}\right)_{\rm SM}=
\frac{G_F^2}{\pi}M_NC_V^2\left(1-\frac{T_N}{T_N^{\rm max}}\right)%\left(1-\frac{T_N}{2E_\nu}-\frac{MT_N}{2E_\nu^2}\right)
\end{equation}
is the SM cross section due to weak interaction~\cite{Drukier:1983gj}, with 
$$
T_N^{\rm max}=\frac{2E_\nu^2}{2E_\nu+M_N}.
$$
The second, helicity-flipping component is due to the magnetic moment only and is given
by~\cite{Vogel:1989iv}
\begin{equation}
\label{NMM}
\left(\frac{d\sigma_{\nu_{e}N}}{dT_{N}}\right)_{\rm mag}=4\pi
\alpha\mgm_{\nu_e}^2\,\frac{Z^2}{T_N}\left(1-\frac{T_N}{E_\nu}+\frac{T_N^2}{4E_\nu^2}\right).
\end{equation}
%

%Clearly, any deviation of the measured cross section of the
%process under discussion from the well-known Standard Model
%value~(\ref{SM}) will provide a signature of the physics beyond
%the Standard Model (see
%also~\cite{Scholberg:2005qs,Barranco:2005yy,Barranco:2007tz,Davidson:2003ha}).
Formulas~(\ref{h.-c.}) and~(\ref{NMM}) describe a deviation
from the well-known SM value~(\ref{SM}) due to neutrino electromagnetic interactions. Two important features should be noted. First, the contributions from the neutrino millicharge and charge radius interfere with that from the weak interaction, while the neutrino magnetic moment contributes separately. Second, the roles of the neutrino millicharge and magnetic moment grow with lowering the energy transfer $T_N$, in particular, when $T_N\to0$ the $\chg_{\nu_e}$ contribution behaves as $\propto1/T_N^2$ and the $\mgm_{\nu_e}$ contribution as $\propto1/T_N$.

It can be noted that the characteristic energy scale where $(d\sigma_{\nu_{e}N}/dT_N)_{\rm mag}$ exceeds $(d\sigma_{\nu_e N}/dT_N)_{\rm SM}$,
\begin{equation}
T_{N} \lesssim \frac{\pi^2\alpha^2}{G_F^2M_Nm_e^2}\left(\frac{Z}{C_V}\right)^2
\left(\frac{\mgm_{\nu_\afl}}{\bmag}\right)^2,
\end{equation}
appears to be by orders of magnitude smaller when compared to that for the elastic neutrino-electron scattering~(\ref{nu-e_mag_SM}).
%%
%\section{Conclusion}
%\label{conclusions}
%%
%The above theoretical findings strongly support the upper limit
%$\mu_\nu < 3.2 \times 10^{-11}\mu_B$ recently reported by the
%GEMMA collaboration~\cite{ge2,ge3}. This bound obtained in
%terrestrial experiments with reactor (anti)neutrinos is only by an
%order of magnitude weaker than the most stringent astrophysical
%constraint $\mu_\nu < 3 \times 10^{-12}\mu_B$~\cite{raffelt90}. A
%general and model-independent upper bound on the Dirac NMM, that
%can be generated by an effective theory beyond the SM, is
%$\mu_\nu\leq10^{-14}\mu_B$~\cite{bell05} (the limit in the
%Majorana case is much weaker). Thus, the searches for NMM are
%close to the territory where new physics can reveal itself.

%
\ack
We are thankful to Nicolao Fornengo and Carlo Giunti for
the kind invitation to participate in the 14th International Conference on Topics in Astroparticle
and Underground Physics. This work was supported by RFBR grant nos. 14-22-03043 ofi\_m, 15-52-53112 GFEN\_a, and 16-02-01023 A. One of the authors (K.A.K.) also acknowledges support from the RFBR under grant no. 14-01-00420 A.
%One of the authors (A.I.S.) is thankful to Lothar Oberauer, Georg
%Raffelt and Robert Wagner for the kind invitation to participate
%in the 12th International Conference on Topics in Astroparticle
%and Underground Physics. 

%
\section*{References}


\begin{thebibliography}{99}
%
\bibitem{Beringer:1900zz} 
Beringer J et al 2012 \emph{Phys. Rev. D} {\bf 86} 010001
%
\bibitem{Giunti:2007ry}
Giunti C and Kim C W 2007 \emph{Fundamentals of Neutrino
Physics and Astrophysics} (New York: Oxford University Press)
%
\bibitem{Bilenky:2010zza}
Bilenky S 2010 \emph{Lect. Notes Phys.} {\bf 817} 1
%
\bibitem{Xing:2011zza}
Xing Z-Z and Zhou S 2011 \emph{Neutrinos in Particle Physics,
Astronomy and Cosmology} (Hangzhou: Zhejiang University
Press).
%
\bibitem{GonzalezGarcia:2012sz}
Gonzalez-Garcia M, Maltoni M, Salvado J and Schwetz T 2012
\emph{JHEP} {\bf 1212} 123
%
\bibitem{Bellini:2013wra}
Bellini G, Ludhova L, Ranucci G and Villante F 2014 \emph{Adv. High Energy Phys.} {\bf 2014} 191960
%
\bibitem{Pontecorvo:1957cp}
Pontecorvo B 1957 \emph{Sov. Phys. JETP} {\bf 6} 429
%
\bibitem{Pontecorvo:1957qd}
Pontecorvo B 1958 \emph{Sov. Phys. JETP} {\bf 7} 172
%
\bibitem{Maki:1962mu}
Maki Z, Nakagawa M and Sakata S 1962 \emph{Prog. Theor.
Phys.} {\bf 28} 870
%
\bibitem{Pontecorvo1968}
Pontecorvo B 1968 \emph{Sov. Phys. JETP} {\bf 26} 984
%
\bibitem{Ramond:1999vh}
Ramond P 1999 \emph{Journeys beyond the Standard Model}
(Cambridge, MA: Perseus Books)
%
\bibitem{Mohapatra:2004}
Mohapatra R N and Pal P B 2004 \emph{Massive Neutrinos
in Physics and Astrophysics} (Singapore: World Scientific)
%
\bibitem{Giunti_RMP2015}
Giunti C and Studenikin A 2015 \emph{Rev. Mod. Phys.} {\bf 87} 531
%
\bibitem{Fujikawa:1980yx}
  Fujikawa K and Shrock R 1980
  %``The Magnetic Moment Of A Massive Neutrino And Neutrino Spin Rotation,''
  \emph{Phys. Rev. Lett.}  {\bf 45} 963
%
\bibitem{Beda:2012zz}
Beda A et al 2012 \emph{Adv. High Energy Phys.} {\bf 2012} 350150
%
\bibitem{Raffelt:1990pj}
Raffelt G 1990 \emph{Phys. Rev. Lett.} {\bf 64} 2856
%
\bibitem{Wong:2005pa}
Wong H T and Li H B 2005 \emph{Mod. Phys. Lett. A} {\bf
20} 1103
%
\bibitem{Balantekin:2006sw}
Balantekin A 2006 \emph{AIP Conf. Proc.} {\bf 847} 128
%
\bibitem{Beda:2007hf}
Beda A G et al 2007 \emph{Phys. Atom. Nucl.} {\bf 70} 1873 
%(\emph{Preprint} hep-ex/0705.4576)
%
\bibitem{Giunti:2008ve}
Giunti C and Studenikin A 2009
\emph{Phys. Atom. Nucl.}  {\bf 72} 2089 %(\emph{Preprint} hep-ph/0812.3646)
%
\bibitem{Broggini:2012df}
Broggini C, Giunti C and Studenikin A 2012 \emph{Adv. High
Energy Phys.} {\bf 2012} 459526
%
\bibitem{Carlson:1932rk}
Carlson J and Oppenheimer J 1932 \emph{Phys. Rev.} {\bf 41} 763
%
\bibitem{Bethe:1935cp}
Bethe H 1935 \emph{Math. Proc. Cambridge Philos. Soc.} {\bf 31} 108
%
\bibitem{Kyuldjiev:1984kz}
Kyuldjiev A V 1984 \emph{Nucl. Phys. B} {\bf 243} 387
%
\bibitem{Domogatsky:1971tu}
Domogatsky G and Nadezhin D 1970 \emph{Yad. Fiz.} {\bf 12} 1233
%
\bibitem{Cowan:1954pq}
Cowan C, Reines  F and Harrison F 1954 \emph{Phys. Rev.} {\bf 96} 1294
%
\bibitem{Cowan:1957pp}
Cowan C and Reines F 1957 \emph{Phys. Rev.} {\bf 107} 528
%
\bibitem{Vogel:1989iv}
Vogel P and Engel J 1989 \emph{Phys. Rev. D} {\bf 39} 3378
%
\bibitem{Kouzakov:2014ahe}
Kouzakov K A and Studenikin A I 2014 \emph{Adv. High Energy
Phys.} {\bf 2014} 569409
%
\bibitem{Freedman:1973yd}
Freedman D Z 1974 \emph{Phys. Rev. D} {\bf 9} 1389
%
\bibitem{Wong:2011zzb}
Wong H T 2011 \emph{Int. J. Mod. Phys. D} {\bf 20} 1463
%
\bibitem{Li:2013fla}
Li H et al 2013 \emph{Phys. Rev. Lett.} {\bf 110} 261301
%
\bibitem{Li:2013ewa}
Li H et al 2014 \emph{Astropart. Phys.} {\bf 56} 1
%
\bibitem{Drukier:1984}
Drukier A and Stodolsky L 1984 \emph{Phys. Rev. D} {\bf 30} 2295
%
\bibitem{Kouzakov:2010tx}
Kouzakov K A and Studenikin A I 2011 \emph{Phys. Lett. B} {\bf 696} 252
%
\bibitem{Kouzakov:2011ig}
Kouzakov K A and Studenikin A I 2011 \emph{Nucl. Phys. B (Proc. Suppl.)} {\bf 217} 353
%
\bibitem{Kouzakov:2011ka}
Kouzakov K A, Studenikin A I and Voloshin M B 2011 \emph{Phys. Rev D} {\bf 83} 113001
%
\bibitem{Kouzakov:2011vx}
Kouzakov K A, Studenikin A I and Voloshin M B
2011 \emph{JETP Lett.} {\bf 93} 623
%
\bibitem{Kouzakov:2011uq}
Kouzakov K A, Studenikin A I and Voloshin M B 2012 \emph{J. Phys.: Conf. Series} {\bf 375} 042045
%
\bibitem{kopeikin97}
Kopeikin V I, Mikaelyan L A, Sinev V V and Fayans S A 1997 \emph{Phys. At. Nucl.} {\bf 60} 1859
%
\bibitem{Kouzakov:2014pepanlett}
Kouzakov K A and Studenikin A I 2014 \emph{Phys. Part. Nucl. Lett.} {\bf 11} 458
%
\bibitem{Chen:2014plb}
Chen J-W et al 2014 \emph{Phys. Lett. B} {\bf 731} 159
%
\bibitem{Kouzakov:2016ichep}
Kouzakov K A and Studenikin A I 2014 Theory of ionizing neutrino-atom collisions: The role of
atomic recoil \emph{Preprint hep-ph/1412.7061}
%
\bibitem{Kouzakov:2015nppp}
Kouzakov K A and Studenikin A I 2015 \emph{Nucl. Part. Phys. Proc.} {\bf 265-266} 323
%
\bibitem{Drukier:1983gj}
Drukier A and Stodolsky L 1984 \emph{Phys. Rev. D} {\bf 30} 2295
%

%%
%\bibitem{studenikin09}Studenikin A I 2009 \emph{Nucl. Phys. B (Proc. Suppl.)} {\bf 188}
%220
%%
  %
%\bibitem{tx}
  %Wong H T et al 2007
  %%``Search of Neutrino Magnetic Moments with a High-Purity germanium Detector
  %%at the Kuo-Sheng Nuclear Power Station,''
  %\emph{Phys. Rev. D} {\bf 75} 012001 (\emph{Preprint} hep-ex/0605006).
%\bibitem{ge2}
 %Beda A G et al 2010
  %%``GEMMA experiment: Three years of the search for the neutrino magnetic
  %%moment,''
  %\emph{Phys. Part. Nucl. Lett.}  {\bf 7} 406 (\emph{Preprint} hep-ex/0906.1926)
  %%%CITATION = 00438,7,406;%%
%\bibitem{ge3}
 %Beda A G et al 2010 \emph{Preprint} hep-ex/1005.2736
  %%``Upper limit on the neutrino magnetic moment from three years of data from
  %%the GEMMA spectrometer,''
%%
%\bibitem{shrock82}Shrock R E 1982 \emph{Nucl. Phys. B} {\bf 206} 359
%%
%\bibitem{kayser82}Kayser B 1982 \emph{Phys. Rev. D} {\bf 26} 1662
%%
%\bibitem{kayser84}Kayser B 1984 \emph{Phys. Rev. D} {\bf 30} 1023
%%
%\bibitem{nieves82}Nieves J F 1982 \emph{Phys. Rev. D} {\bf 26} 3152
%%
%\bibitem{bell05}Bell N et al 2005 \emph{Phys. Rev. Lett.} {\bf 95} 151802
%
\end{thebibliography}
\end{document}